# Modelling Fatigue Behaviours and Lifetimes of Novel GLARE Laminates under Random Loading Spectrum

Zheng-Qiang Cheng[a,b], Wei Tan[c], Jun-Jiang Xiong[a,*], Er-Ming He[d], Tao-Huan Xiong[a], Ying-Peng Wang[a]

[a] School of Transportation Science and Engineering, Beihang University, Beijing 100191, People's Republic of China (*Corresponding author. E-mail address: jjxiong@buaa.edu.cn)

[b] School of Mechanics and Aerospace Engineering, Southwest Jiaotong University, Chengdu 610031, People's Republic of China

[c] School of Engineering and Materials Science, Queen Mary University of London, London E1 4NS, United Kingdom

[d] School of Aeronautics, Northwestern Polytechnical University, Xi'an 710072, People's Republic of China

**Abstract**: This paper aims to experimentally and numerically probe fatigue behaviours and lifetimes of novel GLARE (glass laminate aluminium reinforced epoxy) laminates under random loading spectrum. A mixed algorithm based on fatigue damage concepts of three-phase materials was proposed for modelling progressive fatigue damage mechanisms and fatigue life of fibre metal laminates (FML) under random loading spectrum. To validate the proposed modelling algorithm, fatigue tests were conducted on the GLARE 2/1 and GLARE 3/2 laminates subjected to random loading spectrum, and fatigue mechanisms were discussed by using scanning electron microscope (SEM) analysis. It is shown that predominant fatigue failure of the GLARE laminate depends on the reference load level of random loading spectrum. Specifically, dominant fatigue failure of the GLARE laminate is dependent on fatigue strength of fibre layer at a high reference load level, but metal layer at a low reference load level. Numerical predictions agree well with experimental results, demonstrating that the proposed mixed modelling algorithm can effectively simulate fatigue behaviours and lives of the GLARE laminate under random loading spectrum.

**Nomenclature**

| | |
|---|---|
| $a$ | crack length |
| $C, m, \alpha, \beta$ | material constants in fatigue initiation model for metal |
| $C_1, m_1, m_2$ | material constants in fatigue crack growth model for metal |
| $C_3, m_3$ | material constants in fatigue delamination growth model on fibre-metal interface |
| $d$ | damage variable |
| $d_t$ | total accumulated delamination damage |
| $d^s$ | static damage variable in delamination growth model |
| $d^f$ | fatigue damage variable in delamination growth model |
| $da/dN$ | fatigue crack growth rate |
| $db/dN$ | fatigue delamination growth rate |
| $E$ | Young's modulus |
| $E_1$ | Young's modulus of metal layer |
| $E_2$ | Young's modulus of fibre layer |
| $E'$ | damaged Young's modulus |
| $F$ | external load |
| $G$ | shear modulus |
| $G'$ | damaged shear modulus |
| $G_{IC}, G_{IIC}$ | critical energy release rate for modes I and II delamination growth |
| $G_{IImax}$ | maximum energy release rate for mode II delamination growth |
| $H, p, q$ | material constants in multiaxial strength degradation model |
| $k_I, k_{II}$ | normal and shear penalty stiffness |
| $K_t$ | stress concentration factor |
| $l$ | total number of fatigue cycles in random loading spectrum |
| $L_e$ | length of interface element in the direction of delamination growth |
| $n$ | number of cyclic stress cycles |
| $N$ | number of cyclic stress cycles to fatigue failure |
| $r$ | arbitrary stress ratio |
| $r_0$ | specific stress ratio |
| $r_{eff}$ | effective stress ratio |
| $r_{so}$ | overload shut-off ratio |
| $R(n)$ | fatigue residual strength at cyclic stress cycles $n$ |
| $S$ | nominal stress |
| $S_a$ | amplitude of nominal stress |
| $S_{L,max}$ | maximum local von Mises stress |
| $S_{max}$ | maximum value of nominal stress |
| $S_{max,r}$ | maximum value of nominal stress at arbitrary stress ratio |
| $S_{max,eff}$ | maximum effective nominal stress |

| | | |
|---|---|---|
| $S_{max,OL}$ | maximum nominal stress for overload stress cycle | |
| $S_{min,r}$ | minimum value of nominal stress at arbitrary stress ratio | |
| $S_{r_0}$ | maximum absolute value of fatigue stress at specific stress ratio | |
| $S_0$ | fatigue endurance limit | |
| $t_1$ | thickness of metal layer | |
| $V_1$ | metal volume fraction | |
| $w$ | width of specimen | |
| $X$ | static strength | |
| $X_I$ | normal strength of fibre-metal interface | |
| $X_{II}$ | shear strength of fibre-metal interface | |
| $Y$ | shape function | |
| $z_{OL}$ | size of overload zone | |
| $\varepsilon_p$ | plastic strain | |
| $\eta$ | B-K mixed mode power | |
| $\nu$ | Poisson's ratio | |
| $\nu'$ | damaged Poisson's ratio | |
| $\rho$ | density | |
| $\sigma$ | stress | |
| $\sigma_{II}$ | traction stress for mode II delamination | |
| $\sigma_s$ | yielding strength | |
| $\sigma_u$ | ultimate tension strength | |
| $\Delta a$ | crack growth increment | |
| $\Delta D$ | fatigue damage increment | |
| $\Delta a'$ | crack growth increment through the overload zone | |
| $\Delta D'$ | fatigue damage increment through the overload zone | |
| $\Delta K$ | stress intensity factor range | |
| $\Delta K_{th}$ | crack growth threshold value | |
| $\Delta X$ | strength reduction | |
| $\delta_{II}$ | separation displacement for mode II delamination | |
| $\xi$ | stress triaxiality | |
| $\bar{u}_f^{pl}$ | equivalent plastic displacement at failure | |
| $\dot{\bar{\varepsilon}}^{pl}$ | equivalent plastic strain rate | |
| $\bar{\varepsilon}_o^{pl}$ | equivalent plastic strain at damage initiation | |

**Notation and Acronym**

| t | tension | 1t | tension along longitudinal direction |
|---|---|---|---|
| c | compression | 1c | compression long longitudinal direction |
| 11 | longitudinal direction | 2t | tension along transverse direction |
| 12 | longitudinal-transverse direction | 2c | compression along transverse direction |
| 13 | longitudinal-through thickness direction | FE | finite element |
| 22 | transverse direction | FML | fibre metal laminate |
| 23 | transverse-through thickness direction | GLARE | glass laminate aluminium reinforced epoxy |

# 1 Introduction

Fibre metal laminates (FMLs) are hybrid composite materials composed of alternating metal sheets and fibre-reinforced polymer matrix plies[1]. In comparison to monolithic metals and conventional composites, the FMLs have superior fatigue and damage tolerance behaviours, better resistances to impact, corrosion and flame[2]. Glass laminate aluminium reinforced epoxy (GLARE) as the second generation of the FML has been widely applied in aircraft parts, such as fuselage skins, vertical or horizontal tail leading edges[3]. Due to the complex failure mechanisms of the FML arising from multiple phase material characteristics, its structural integrity in service remains one of the major challenges to the aviation industry[4]. It has been reported that static and fatigue behaviours and failure mechanisms of the FML are significantly affected by numerous factors, such as metal types and volume fraction[5], metal surface treatment technology[6], fibre type[7], stacking sequence[8], environment condition[9], loading type[10]错误!未找到引用源。, loading rate[12], and others.

The interactions of fatigue failure mechanisms in the FML such as fatigue crack initiation and growth, and fibre-metal interface delamination growth are complicated. Previous works of literature manifest that:

- **(i)** Fatigue initiation life of the GLARE laminate is determined only by stress cycles in metal layers[1], and decreases as the off-axis angle increases from 0° to 45°[13]. Besides, the exposure to a combination of moisture and elevated temperature (85% humidity, 70°C, 3000 h) has not affected fatigue initiation behaviours of the GLARE laminates[1], implicating that the GLARE laminates are not sensitive to high temperature and humidity environment.
- **(ii)** Fatigue crack growth behaviours of the FML is dependent on stress intensity factor caused by

far-field applied loading and by fibre bridging mutually[14], and crack growth threshold of the FML is solely related to metal layer[15].

**(iii)** Fatigue delamination threshold on fibre-metal interfaces is governed by epoxy matrix, while fatigue delamination growth rate is dominated by the cohesion of fibre-matrix. Generally, fatigue delamination resistance of the FML reinforced by glass fibres is higher than that reinforced by carbon fibres[16]. Failure mode on the interfaces between metal layers and adjacent intact fibre layers are mainly mode II delamination under fatigue loading[17].

**(iv)** Fatigue delamination growth rate of the GLARE laminate is independent on the load sequences[18]. Moreover, the acceleration effect of crack growth in the GLARE laminate as a result of the underloads is insignificant, while the retardation effect due to the overloads is significant. However, the magnitude of retardation effect is less in the GLARE laminate than that in monolithic aluminium alloys because of the fibre bridging[19].

Although a large number of studies[13]-[19] have been conducted to investigate fatigue failure mechanisms and behaviours of three-phase materials (i.e., fibre layer, metal layer, and matrix layer on fibre-metal interface) in the FML, those studies on interactive fatigue failure mechanisms of three-phase materials are very limited. In addition, current studies mainly focus on probing constant amplitude fatigue behaviours of the FML, but there are few works on fatigue failure mechanisms and fatigue lifetimes of the FML under random loading spectrum.

To predict fatigue failure mechanisms and fatigue life of the FML, two major approaches have been devised. One is the method of combining fatigue and fracture mechanics (or damage mechanics) for single-phase material, which hypotheses that fatigue life of the FML is governed only by the stress cycles in metal layers. The $S-N$ and $da/dN-\Delta K$ curves of monolithic metal are employed to estimate crack initiation and growth lives of metal layers, and then fatigue life of the FML is obtained by summing both lives[10][20][21]. The drawbacks of this method though are its intensity and complexity when the effects of fibre bridging and delamination growth on crack growth behaviours are considered, and it neglects the impact of fibre and matrix failures in fibre layer on fatigue life of the FML. This has led to the development of the second fatigue method for dual-phase materials, in which fatigue life of the FML is dominated mutually by both fatigue behaviours of metal and fibre layers. Fatigue properties of monolithic metal and monolithic composite are adopted to predict fatigue lives of metal and fibre layers in the FML, respectively[2][7][22]. However, this method does not consider the effect of fibre-metal interface delamination yet. Thereby, it is crucial to develop an FML fatigue life prediction approach that deals with the interactive fatigue failure mechanisms of three-phase materials.

In view of this, the aforementioned research gaps are investigated in this paper. The main novel contributions herein are: **(i)** A mixed algorithm based on fatigue damage concepts of three-phase materials (i.e., fibre layer, metal layer, and matrix layer on fibre-metal interface) is proposed for modelling progressive fatigue damage mechanisms and fatigue life of the FML under random loading spectrum. **(ii)** Fatigue tests are performed on two kinds of novel GLARE laminates subjected to the random loading spectrum and fatigue failure mechanisms are discussed by using scanning electron microscope (SEM) analysis. **(iii)** The numerical predictions from this work agree well with experimental data, revealing the complex interaction of various fatigue failure mechanisms. Our work opens a new avenue to numerically predict fatigue life of the FML under random loading spectrum.

This paper is organised as follows: The introduction part presents research gaps about fatigue behaviours and lifetimes of FML. Section 2 illustrates the mixed algorithm based on fatigue damage concepts of three-phase materials in detail. Section 3 gives the static and fatigue test results including SEM analysis for depicting the fatigue failure mechanisms of FML. Section 4 shows numerical analysis results to verify the developed mixed algorithm. Section 5 summarises experimental and numerical results.

## 2 Mixed algorithm based on fatigue damage concepts of three-phase materials in the FML under random loading spectrum

The FML always consists of the fibre layer, metal layer, and matrix layer on fibre-metal interface. Mechanical properties and failure modes of three-phase materials are significantly different, and failure mechanisms of three-phase materials are interactive under fatigue loading. Hence, in order to model fatigue failure mechanisms and fatigue life of the FML under random loading spectrum, it is essential to analyse fatigue damage behaviours of three-phase materials and to establish a progressive fatigue damage algorithm considering the effect of load sequence and the interactive fatigue damage mechanisms of three-phase materials.

### 2.1 Fatigue model of fibre layer

Strength and stiffness properties of fibre layer in longitudinal, transverse, in-plane shear and out-plane shear directions could degrade under repeated fatigue loading. Hence, multiaxial strength degradation model and sudden stiffness degradation rule developed by authors' previous work[23]-[25] have been employed to characterise the strength and stiffness degradation behaviours of fibre layer, respectively. In brief, the multiaxial strength degradation model is based on the residual strength model which characterise the gradual strength degradation of composites under fatigue cycles[23]. For the sudden stiffness degradation rule, the stiffness is assumed to be unchanged before fatigue failure criteria are triggered, otherwise, the stiffness properties of failed composites are degraded to nearly zero[24]. Those formulations are as follow:

$$\begin{cases} \Delta X_{it}(n) = \left[ \left( \Delta X_{it}(n-1) \right)^{q_{it}} + H_{it}^{-1} \left( S_{r_0} - S_{0,it} \right)^{-p_{it}} \right]^{\frac{1}{q_{it}}} & (i=1,2) \\ \Delta X_{ic}(n) = \left[ \left( \Delta X_{ic}(n-1) \right)^{q_{ic}} + H_{ic}^{-1} \left( S_{r_0} - S_{0,ic} \right)^{-p_{ic}} \right]^{\frac{1}{q_{ic}}} & (i=1,2) \\ \Delta X_{ij}(n) = \left[ \left( \Delta X_{ij}(n-1) \right)^{q_{ij}} + H_{ij}^{-1} \left( S_{r_0} - S_{0,ij} \right)^{-p_{ij}} \right]^{\frac{1}{q_{ij}}} & (i,j=1,2,3, i \neq j) \end{cases} \quad (1)$$

$$\begin{cases} E'_{ii} = (1-d_{ii}) E_{ii} & (i=1,2,3) \\ v'_{ij} = \dfrac{E'_{ii} v_{ij}}{E_{ii}} & (i,j=1,2,3, i \neq j) \\ G'_{ij} = (1-d_{ij}) G_{ij} & (i,j=1,2,3, i \neq j) \end{cases} \quad (2)$$

with

$$\begin{cases} \Delta X_{it}(n) = X_{it} - R_{it}(n) & (i=1,2) \\ \Delta X_{ic}(n) = X_{ic} - R_{ic}(n) & (i=1,2) \\ \Delta X_{ij}(n) = X_{ij} - R_{ij}(n) & (i,j=1,2,3, i \neq j) \end{cases} \quad (3)$$

$$S_{r_0} = \begin{cases} \dfrac{(1-r) X_{it} S_{itmax,r}}{(1-r_0) X_{it} + (r_0 - r) S_{itmax,r}} & (i=1,2), (r_0^2 \leq 1, r^2 \leq 1) \\ \dfrac{(r-1) r_0 X_{ic} |S_{icmin,r}|}{(r_0 -1) r X_{ic} - (r_0 - r) |S_{icmin,r}|} & (i=1,2), (r_0^2 > 1, r^2 > 1) \end{cases} \quad (4)$$

$$S_{ij} = \begin{cases} \dfrac{(1-r) X_{ij} S_{ijmax,r}}{(1-r_0) X_{ij} + (r_0 - r) S_{ijmax,r}} & (i,j=1,2,3, i \neq j), (r_0^2 \leq 1, r^2 \leq 1) \\ \dfrac{(r-1) r_0 X_{ij} |S_{ijmin,r}|}{(r_0 -1) r X_{ij} - (r_0 - r) |S_{ijmin,r}|} & (i,j=1,2,3, i \neq j), (r_0^2 > 1, r^2 > 1) \end{cases} \quad (5)$$

$$\begin{cases} d_{ii} = 1 - (1-d_{ii}^t)(1-d_{ii}^c) & (i=1,2,3) \\ d_{12} = d_{11} \\ d_{13} = d_{11} \\ d_{23} = \max(d_{22}, d_{33}) \end{cases} \quad (6)$$

where:

$X_{it}$, $X_{ic}$, $X_{ij}$ are the static tension, compression and shear strengths of fibre layer, respectively;

$\Delta X_{it}(n)$, $\Delta X_{ic}(n)$, $\Delta X_{ij}(n)$ are the reduction values in tension, compression and shear strengths after $n$ number of fatigue loading cycles, respectively;

$S_{r_0}$ and $S_{ij}$ are the maximum absolute values of fatigue stress at specific stress ratio in normal and shear directions, respectively;

$n$ is the number of fatigue loading cycles;

$R_{it}(n)$, $R_{ic}(n)$, $R_{ij}(n)$ are the residual tension, compression and shear strength after $n$ number of fatigue loading cycles, respectively;

$r$ is the arbitrary stress ratio which equals to the ratio of minimum and maximum stress of a stress cycle in random loading spectrum;

$r_0$ is the specific stress ratios which means the ratio of minimum and maximum stress of a stress cycle under experimental conditions;

$H_{it}, H_{ic}, H_{ij}, p_{it}, p_{ic}, p_{ij}, q_{it}, q_{ic}, q_{ij}, S_{0,it}, S_{0,ic}, S_{0,ij}$ are the constants in multiaxial strength degradation model and can be determined by using best fitting method[23];

$E'_{ii}, E_{ii}, G'_{ij}, G_{ij}, v'_{ij}, v_{ij}$ are the damaged and undamaged Young's modulus, shear modulus and Poisson's ratio, respectively;

$d$ is the damage variable, and the value of $d$ is assumed to be zero before fatigue failure criteria are triggered, otherwise, it is valued as 0.99 according to practical sudden stiffness degradation rule[24].

Notably, the effects of stress ratio and load sequence under random loading spectrum are taken into account in multiaxial strength degradation model (1) by the cycle-by-cycle calculation[23].

The Olmedo failure criteria[26] have been successfully employed to identify four typical failure modes (including fibre tension and compression failures, matrix tension and compression failures) in composites under static loading. However, it is unavailable to predict fatigue failure modes of composites because it neglects gradual strength degradation under fatigue loading. For this reason, material's strengths in the Olmedo's failure criteria are replaced by multiaxial fatigue residual strengths to derive fatigue failure criteria[25] (shown in Table 1). The developed fatigue failure criteria have been used to identify potential fatigue failure modes for fibre layer in the FML under random loading spectrum.

Table 1　Fatigue failure criteria of fibre layer.

| Fibre tension fatigue failure | $\left(\dfrac{\sigma_{11}}{X_{1t}-\Delta X_{1t}(n)}\right)^2+\left(\dfrac{\sigma_{12}}{X_{12}-\Delta X_{12}(n)}\right)^2+\left(\dfrac{\sigma_{13}}{X_{13}-\Delta X_{13}(n)}\right)^2\geq 1$ |
|---|---|
| Fibre compression fatigue failure | $\left(\dfrac{\sigma_{11}}{X_{1c}-\Delta X_{1c}(n)}\right)^2\geq 1$ |
| Matrix tension fatigue failure | $\left(\dfrac{\sigma_{22}}{X_{2t}-\Delta X_{2t}(n)}\right)^2+\left(\dfrac{\sigma_{12}}{X_{12}-\Delta X_{12}(n)}\right)^2+\left(\dfrac{\sigma_{23}}{X_{23}-\Delta X_{23}(n)}\right)^2\geq 1$ |
| Matrix compression fatigue failure | $\left(\dfrac{\sigma_{22}}{X_{2c}-\Delta X_{2c}(n)}\right)^2+\left(\dfrac{\sigma_{12}}{X_{12}-\Delta X_{12}(n)}\right)^2+\left(\dfrac{\sigma_{23}}{X_{23}-\Delta X_{23}(n)}\right)^2\geq 1$ |

## 2.2 Fatigue model of metal layer

Fatigue crack initiation and growth of metal layer in the FML can be respectively characterised as[27][28]

$$\left\{\frac{2\sigma_s S_a(1-r)}{\sigma_s(1-r)(1-r_0)+2S_a(r_0-r)}-\left[1+\alpha\left(1-K_t^\beta\right)\right]S_0\right\}^m N = C \tag{7}$$

$$\frac{da}{dN}=C_1(\Delta K-\Delta K_{th})^{m_1}(1-r)^{m_2} \tag{8}$$

with

$$K_t=\frac{S_{L,max}}{S} \tag{9}$$

$$\Delta K = 2Y(a/w)S_a\sqrt{\pi a} \tag{10}$$

where:

$\alpha$, $\beta$, $C$ and $m$ are the material's constants in fatigue initiation model of metal;

$S_0$ is the fatigue endurance limit; the parameters of $\alpha$, $\beta$, $C$, $m$ and $S_0$ are estimated from the data from constant amplitude fatigue tests at various stress concentration factors by the Least Squares Fitting method[27];

$N$ is the number of cyclic stress cycles to fatigue failure;

$S_a$ is the amplitude of nominal stress in metal layer;

$\sigma_s$ is the yielding strength of metal layer;

$C_1$, $m_1$ and $m_2$ are the material's constants in fatigue crack growth model of metal, and are determined by constant amplitude fatigue crack growth tests at various stress ratios[28];

$da/dN$ is the fatigue crack growth rate;

$\Delta K$ and $\Delta K_{th}$ are the stress intensity factor range and crack growth threshold, respectively;

$K_t$ is the stress concentration factor;

$S_{L,max}$ and $S$ are the maximum local von Mises stress and nominal stress in metal layer, respectively;

$Y(a/w)$ is the shape function;

$a$ is the crack length;

$w$ is the width of the specimen.

Damage variable $D$ is introduced to characterise fatigue damage of metal layer in the FML. Based on the Miner's linear cumulative damage theory and Eq. (7), it is possible to deduce fatigue damage increment $\Delta D_i$ for the $i$-th stress cycle in random loading spectrum, that is

$$\Delta D_i = \frac{1}{N_i} = \frac{1}{C} \left\{ \frac{2\sigma_s S_{a,i}(1-r_i)}{\sigma_s(1-r_i)(1-r_0) + 2S_{a,i}(r_0-r_i)} - \left[1+\alpha\left(1-K_{t,i}^\beta\right)\right]S_0 \right\}^m \tag{11}$$

Similarly, according to Eq. (8), fatigue crack growth increment $\Delta a_i$ for the $i$-th stress cycle in random loading spectrum can be shown to be

$$\Delta a_i = C_1 \left(\Delta K_i - \Delta K_{th}\right)^{m_1} \left(1-r_i\right)^{m_2} \tag{12}$$

Literature [28][29] manifests that load sequence has a remarkable effect on fatigue life of the metal under random loading spectrum. Based on the authors' previous model[28][29] which considers the effect of load sequence, the concept of an effective stress ratio $r_{eff}$ is introduced into Eqs. (11) and (12) to deduce damage increment $\Delta D_i$ and $\Delta a_i$ for the $i$-th stress cycle in random loading spectrum, which is argued to depict the load sequence effect on fatigue damage and life of metal layer, namely

$$\Delta D_i = \frac{1}{C}\left\{\frac{2\sigma_s S_{a,i}(1-r_{\text{eff},i})}{\sigma_s(1-r_{\text{eff},i})(1-r_0)+2S_{a,i}(r_0-r_{\text{eff},i})} - \left[1+\alpha(1-K_{t,i}^{\beta})\right]S_0\right\}^m \tag{13}$$

$$\Delta a_i = C_1(\Delta K_i - \Delta K_{\text{th}})^{m_1}(1-r_{\text{eff},i})^{m_2} \tag{14}$$

with

$$r_{\text{eff}} = 1 - \frac{2S_a}{S_{\text{max,eff}}} \tag{15}$$

$$S_{\text{max,eff}} = S_{\text{max}} - \frac{S_{\text{max,OL}}-S_0}{(r_{\text{so}}-1)S_{\text{max,OL}}}\left(S_{\text{max,OL}}\sqrt{1-\frac{\Delta\Phi'}{z_{\text{OL}}}}-S_{\text{max}}\right) \tag{16}$$

$$z_{\text{OL}} = \frac{1}{2}\left(\frac{S_{\text{max,OL}}\sqrt{\Phi}}{\sigma_s}\right)^2 \tag{17}$$

$$\Phi = \begin{cases} D, & \text{crack initiation} \\ a, & \text{crack growth} \end{cases} \tag{18}$$

where:

$S_{\text{max,eff}}$ is the maximum effective nominal stress;

$S_{\text{max,OL}}$ is the maximum nominal stress for overload stress cycle;

$r_{\text{so}}$ is the overload shutoff ratio;

$\Delta D'$ and $\Delta a'$ are the fatigue damage and crack growth increments through the overload zone, respectively;

$z_{\text{OL}}$ is the size of overload zone.

Eq. (13) is employed to calculate fatigue damage increment for each stress cycle in random loading spectrum, and the cumulative fatigue damage can be obtained with such a cycle-by-cycle accumulation calculation. If fatigue failure criterion of the metal ($D \geq 1$) is triggered in current fatigue cycle, fatigue crack initiation occurs in metal layer. After fatigue crack initiation reaches a certain length, fatigue crack growth increment for each stress cycle in random loading spectrum is then calculated according to Eq. (14). Similarly, current crack length can be obtained with such a cycle-by-cycle accumulation calculation, and metal layer fractures once current crack length is greater than critical crack length.

## 2.3 Delamination growth model of matrix layer on fibre-metal interface

Delamination growth behaviours of matrix layer on fibre-metal interfaces in the FML could cause stress redistribution in fibre and metal layers, which has a significant effect on fatigue life of the FML[30]. Traditional cohesive zone model (CZM) is extended to fatigue CZM for capturing delamination growth of matrix layer on fibre-metal interface under fatigue loading. It is worth noting that predominant fatigue mechanism of matrix layer on fibre-metal interface in the FML under fatigue loading is mode II delamination[17], so only mode II delamination growth is considered in this work. Fatigue delamination growth rate can be described by modified Paris law[31]:

$$\frac{db}{dN} = C_3 \left[ \frac{G_{II\max}}{G_{IIC}} (1-r^2) \right]^{m_3} \quad (19)$$

where:

$db/dN$ is the fatigue delamination growth rate;

$C_3$ and $m_3$ are the material's constants, and can be determined by constant amplitude end-notched flexure fatigue tests;

$G_{IIC}$ and $G_{II\max}$ are the critical energy release rate and maximum energy release rate for mode II delamination growth, respectively.

From the integration of traction versus displacement history, $G_{II\max}$ can be extracted as

$$G_{II\max}(n) = \sum_{k=1}^{n} \left[ \frac{\sigma_{II}(k) + \sigma_{II}(k-1)}{2} \right] \left[ \delta_{II}(k) - \delta_{II}(k-1) \right] \quad (20)$$

where $\sigma_{II}$ and $\delta_{II}$ are the shear stress and separation displacement for mode II delamination, respectively.

According to continuum damage mechanics theory, fatigue delamination growth rate obtained from Eq. (19) in each cohesive element needs to be further converted to fatigue damage variable $d^f$. Moreover, fatigue damage is accumulated only within the cohesive zone where cohesive elements have exceeded their linear-elastic range and experience irreversible deformation[32]. it has been reported that the computation of accumulative fatigue damage for every cohesive element within the cohesive zone leads to a considerably overestimated delamination growth rate[31]. To address this issue, previous works[31][33] confine fatigue damage accumulation to only the delamination-tip elements within the cohesive zone and proposed a delamination-tip tracking algorithm to identify and track delamination-tip elements. One shortcoming of this method is its intensity and complexity of computation. In fact, fatigue cohesive zone coincides with the region in which traditional static damage variable $d^s$ is greater than zero[34]. It seems practical and convenient to set a threshold of static damage variable to identify and focus on the delamination-tip elements within the cohesive zone. Based on the authors' trial simulation results, the threshold of static damage variable valued as 0.95 can effectively capture the delamination-tip elements within the cohesive zone. Therefore, the threshold of static damage variable is reasonably assumed as 0.95 in this work.

According to Eq. (19), it is possible to calculate the number of stress cycles corresponding to a delamination growth length $L_e$ as

$$N_e = L_e \left( \frac{db}{dN} \right)^{-1} \tag{21}$$

where $L_e$ is the effective element length associated with a single cohesive integration point in delamination growth direction.

fatigue damage increment $\Delta d_i^f$ is defined as[31]

$$\Delta d_i^f = \frac{1-d^s}{N_e} \tag{22}$$

and accumulated fatigue damage can be obtained by

$$d^f = \sum_i^l \Delta d_i^f \tag{23}$$

where $l$ is the total number of fatigue cycles in random loading spectrum.

As a result, total accumulated damage within the cohesive zone becomes

$$d_t = d^s + d^f \tag{24}$$

Once total accumulated delamination damage $d_t$ is greater than or equals one, the cohesive element fails and the corresponding number of stress cycles is fatigue life of delamination growth.

## 2.4 Mixed algorithm flowchart

Schematic flowchart of progressive damage analysis for the FML under random loading spectrum is shown in Fig. 1, and the mixed algorithm based on fatigue damage concepts of three-phase materials is written as a main VUMAT subroutine of Abaqus/Explicit software[35]. To address load sequence effect and interactive fatigue failure mechanisms of three-phase materials, main VUMAT subroutine integrates three independent VUMAT subroutine modules of progressive fatigue damage algorithm for fibre layer, metal layer, and matrix layer on fibre-metal interface. Noticeably, although all algorithms are developed on Abaqus/Explicit software, the principle of its algorithm is not limited to this platform, but also applicable to other finite element software (such as Ansys).

Mechanical properties and model parameters of fibre, metal and fibre-metal interface matrix layers are firstly assigned to corresponding constituent materials of the FML as the input data. A random loading spectrum is then applied to the finite element (FE) model by defining the loading amplitude curve. After stress state analysis of the FE model, progressive fatigue damage evaluation of fibre, metal and fibre-metal interface matrix layers is carried out simultaneously in current same fatigue cycle. Once any elements fail, the stress of three-phase material layers will be redistributed. As a result, stress state of the FE model needs to be updated before the next fatigue cycle. With such cycle-by-cycle simulation, fatigue damage of three-phase materials is re-calculated until fatigue failure of the FML happens, and final fatigue failure is marked by the rupture of metal layer on whole cross-section of the specimen. Fatigue life of the FML under random loading spectrum is thus obtained by cumulating fatigue cycle increments until final fatigue failure. Obviously, the effect of load sequence and interactive fatigue failure mechanism of three-phase materials are taken into account in the above progressive fatigue damage analysis.

In detail, for progressive fatigue damage analysis of fibre layer, current strength reduction of fibre layer is calculated by using multiaxial strength degradation model (see Eq. (1)) and fatigue failure criteria are then updated to identify the potential fatigue failure. If fatigue failure of fibre layer happens, stiffness properties of failed elements are degraded according to sudden stiffness reduction rule (see Eq. (2)).

For progressive fatigue damage analysis of metal layer, if equivalent plastic strain of the elements is not greater than zero, nominal stress in metal layer can be obtained as[36]

$$S = \frac{F E_1 V_1}{\left(E_1 V_1 + E_2 (1 - V_1)\right) w t_1} \tag{25}$$

where:

    $V_1$ is the metal volume fraction in the FML;

    $E_1$ and $E_2$ are the Young's moduli of metal and fibre layers, respectively;

    $F$ is the external load;

$t_1$ is the thickness of metal layer.

Substituting Eq. (25) into Eq. (9), stress concentration factor is obtained, and then used to calculate fatigue damage increment in current fatigue cycle according to Eq. (11). Otherwise, if the equivalent plastic strain of elements is greater than zero, nominal stress is approximated to local stress of element, and stress concentration factor is valued as one. Fatigue damage increment in current fatigue cycle is then computed from Eq. (13). Total fatigue damage of metal layer under random loading spectrum is obtained by cumulative fatigue damage increments at failure. If total fatigue damage $D$ meets fatigue failure criterion of the metal (that is, $D \geq 1$), the corresponding elements will be deleted.

Finally, for progressive fatigue damage analysis of fibre-metal interface matrix layer, fatigue delamination damage is accumulated among delamination-tip elements within the cohesive zone. Static delamination damage variable in traditional CZM is firstly implemented to recognise the delamination-tip elements. If static delamination damage variable is greater than or equals to the threshold, the cohesive elements are deemed as the delamination-tip elements within the cohesive zone. For delamination-tip cohesive elements within the cohesive zone, maximum energy release rate for mode II delamination is calculated according to Eq. (20), and then substituted into Eq. (19) to calculate delamination growth rate. At the end, accumulative delamination damage is allowed from Eq. (21) to Eq. (23). If total accumulated delamination damage within delamination-tip cohesive elements (see Eq. (24)) is greater than or equals one, the corresponding cohesive elements will be deleted.

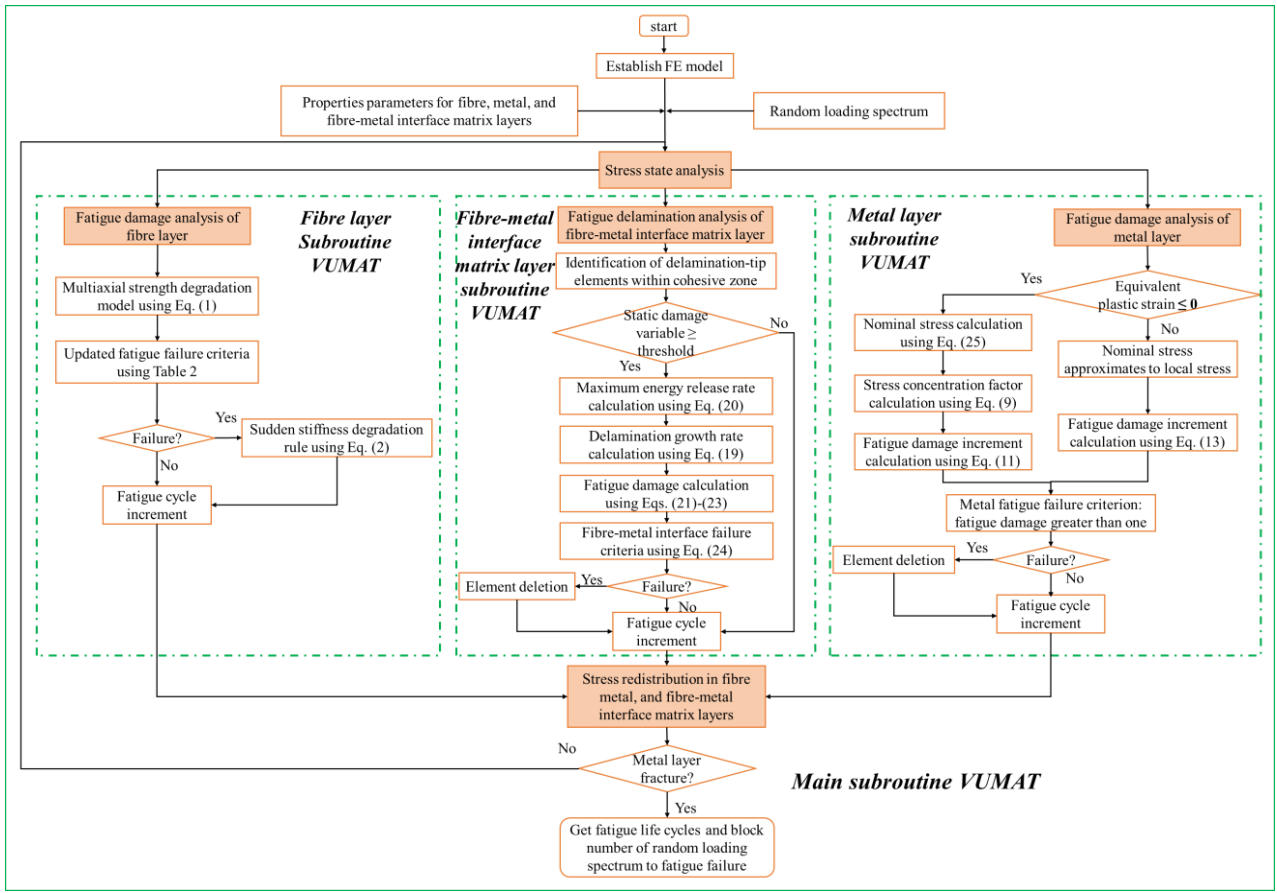

**Fig. 1** Schematic flowchart of progressive fatigue damage analysis for the FML under random loading spectrum.

## 3 Experiment

### 3.1 Materials and specimens

Novel GLARE laminate specimens are made of S4/SY-14 glass fibre prepreg and 2060 Al-Li alloy sheet, and mechanical properties of constituent materials are listed in Table 2. The stacking sequences of two kinds of GLARE laminates are respectively [Al/0/Al] and [Al/0/Al/0/Al], which are marked as the GLARE 2/1 and GLARE 3/2 laminates respectively. "Al" represents the 2060 Al-Li alloy sheet layer, and "0" means the unidirectional longitudinal glass fibre layer. The geometry and dimensions of the GLARE 2/1 and GLARE 3/2 laminate specimens are shown in Fig. 2. The moulding by hot pressing is used to prepare the GLARE laminates. Basic technological processes are the surface treatment of the Al-Li alloy including surface decontamination and chemical oxidation, dry pre-treatment of glass fibre prepreg at 60 °C, alternating laying of the Al-Li alloy and glass fibre prepreg layers, hot pressing curing (temperature 120 °C, stress 6 MPa, time 30 mins), cooling and sampling[10].

**Table 2** Mechanical properties of S4/SY-14 glass fibre lamina and 2060 Al-Li alloy sheet.

| Materials | S4/SY-14 glass fibre lamina[10] | 2060 Al-Li alloy sheet[37] |
| --- | --- | --- |
| Density (g/cm$^3$) | 1.98 | 2.72 |
| Modulus (GPa) | $E_{11}=54.6$; $E_{22}=E_{33}=10.5$; $G_{12}=G_{13}=3.5$; $G_{23}=3.0$ | $E=72.4$ |
| Poisson's ratio | $v_{12}=v_{13}=0.252$; $v_{23}=0.32$ | $v=0.3$ |
| Strength (MPa) | $X_{1t}=2000$; $X_{1c}=1037$; $X_{2t}=X_{3t}=49.8$; $X_{2c}=X_{3c}=149$; $X_{12}=X_{13}=73.7$; $X_{23}=50$ | $\sigma_s=470$ when $\varepsilon_p=0$; $\sigma_u=590$ when $\varepsilon_p=0.105$ |
| Ductile damage parameter | / | $\bar{\varepsilon}_o^{pl}=0.098$; $\xi=0.33$; $\dot{\bar{\varepsilon}}^{pl}=3.14\times10^{-4}$; $\bar{u}_f^{pl}=0.05$ |

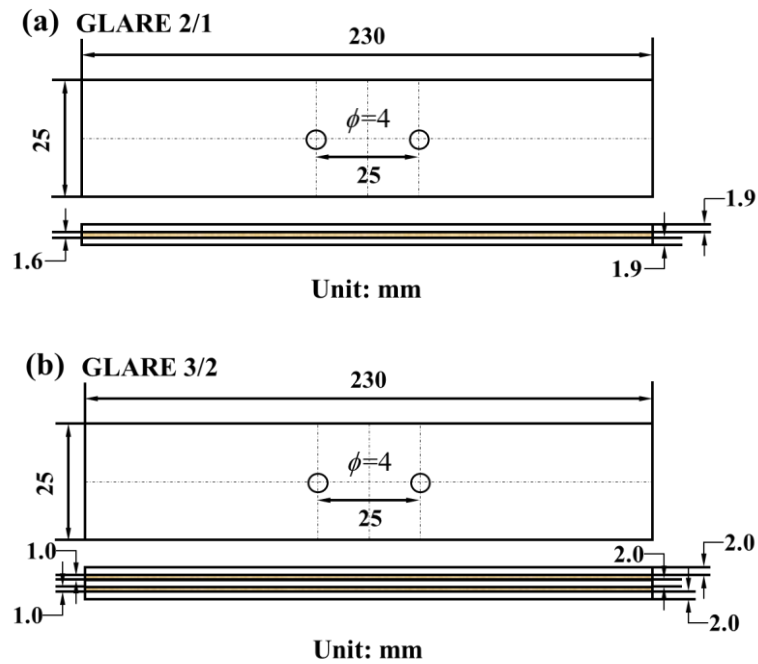

**Fig. 2** Geometry and dimensions: (a) GALRE 2/1 laminate; (b) GALRE 3/2 laminate.

## 3.2 Static and fatigue tests

According to ASTM D3039[38], quasi-static tension tests of the GLARE laminate specimens were carried out on the MTS-810-250kN servo-hydraulic tester at room temperature and moisture by using displacement-control mode, and the loading rate is 2 mm/min. At least two specimens for each type of the GLARE laminate were implemented for tensile tests, and tensile load versus displacement curves automatically were recorded by the test system.

Again, according to ASTM E466[39], fatigue tests of the GLARE laminate specimens under random loading spectrum were conducted on the MTS-880-100kN servo-hydraulic tester at room temperature and moisture by using force-control mode, and the loading frequency was 5 Hz. Fig. 3 shows the random normalised load spectrum, which includes 58442 load cycles counted by the rain-flow counting method. Actual fatigue load is the product of normalised load times reference load level. Taking ultimate tensile loads of the GLAREs as the references, the high, middle and low reference load levels for the GLARE 2/1 laminates are set as 24 kN, 21 kN and 18 kN, respectively, and those for the GLARE 3/2 laminates are 36 kN, 33 kN and 28 kN, respectively. Fatigue failure is defined as the rupture of metal layer in the GLARE laminate specimens. At least four specimens were adopted for each group of fatigue tests to ensure the reliability of the test data.

Fig. 4 shows tensile load versus displacement curves of the GLARE laminates. Fig. 5 presents typical macroscopic failure topologies of the GLARE laminates under random loading spectrum. Table 3 lists the ultimate tensile loads and fatigue lives of the GLARE laminates. The results in Figs. 4 and 5 and Table 3 lead to the following deductions.

**(i)** The curves of quasi-static tensile load versus displacement for the GLARE 2/1 and GLARE 3/2 laminates display a bilinear trend (see Fig. 4). Both the Al-Li alloy and glass fibre layers are within linear elastic regime at initial loading stage, so tensile load increases linearly with the increasing displacement. The initial curve slope of the GLARE 3/2 laminate is bigger than that of the GLARE 2/1 laminate because the GLARE 3/2 laminate has higher metal volume fraction than the GLARE 2/1 laminate. With the further increase in displacement, glass fibre layers still retain elastic deformation, but plastic yielding occurs around circular notch on the Al-Li alloy layer to result in a certain of stiffness reduction. Therefore, global stiffness of the GLARE laminates also decreases gradually, that is, the slope of tensile load versus displacement curve declines gradually. The transition loads for GLARE 2/1 and GLARE 3/2 laminates are approximately 58 kN and 87 kN, respectively. Once plastic yielding appears on the Al-Li alloy layers, major tensile load is carried by glass fibre layers. After that, tensile load versus displacement curve increases linearly with the increase in tensile displacement again. Here, the curve slope of the GLARE 2/1 and GLARE 3/2 laminates are almost same because glass fibre layers dominate the mechanics behaviours. When tensile load reaches, even exceeds ultimate tensile strength of glass fibre layer, the breakage takes place on glass fibre layers, and the Al-Li alloy layers rupture soon. Finally, the GLARE 2/1 and GLARE 3/2 laminates fail at about 88 kN and 130 kN, respectively (see Table 3).

**(ii)** Under random loading spectrum, two typical macroscopic failure topologies occur on notched GLARE 2/1 and GLARE 3/2 laminates. The first one (i.e., Mode I failure) is the rupture of glass fibre and Al-Li alloy layers from a single circular notch, while the second one (i.e., Mode II failure) is the rupture of the Al-Li alloy layer along the width direction of specimen from one circular notch together with fatigue crack growth with a certain length from another circular notch.

**(iii) Again,** two typical macroscopic failure topologies are found to be related to the reference load level of random loading spectrum. In other word, Mode I failure appears on the GLARE laminates under random loading spectrum at high reference load levels of (see Fig. 5(a)), whereas Mode II failure emerges at low reference load levels (see Fig. 5(b)). These results are consistent with the research closures of previous works[7,9]. The reason for this is that at high reference load level of random loading spectrum, plastic stress flows exist in the Al-Li alloy layer and primary bearing ratio shifts from the Al-Li alloy layer into glass fibre layer to carry major fatigue loading. Thereby, glass fibre layers dominate fatigue failure of the GLARE laminate, and the Al-Li alloy layers rapidly fracture followed by fatigue failure of glass fibre layers.

On the other hand, at low reference load level of random loading spectrum, local yielding occurs only around circular notches on the Al-Li alloy layer, causing an insignificant effect on the stiffness of the Al-Li alloy layer. Consequently, the Al-Li alloy layer bears greater fatigue loading than glass fibre layer, and fatigue failure of the GLARE laminates is governed by fatigue strength of the Al-Li alloy layer. It stands to reason that due to the stress concentration around circular notches, fatigue cracks initiate around circular notches, and slowly propagate away from both notches owe to fibre bridging.

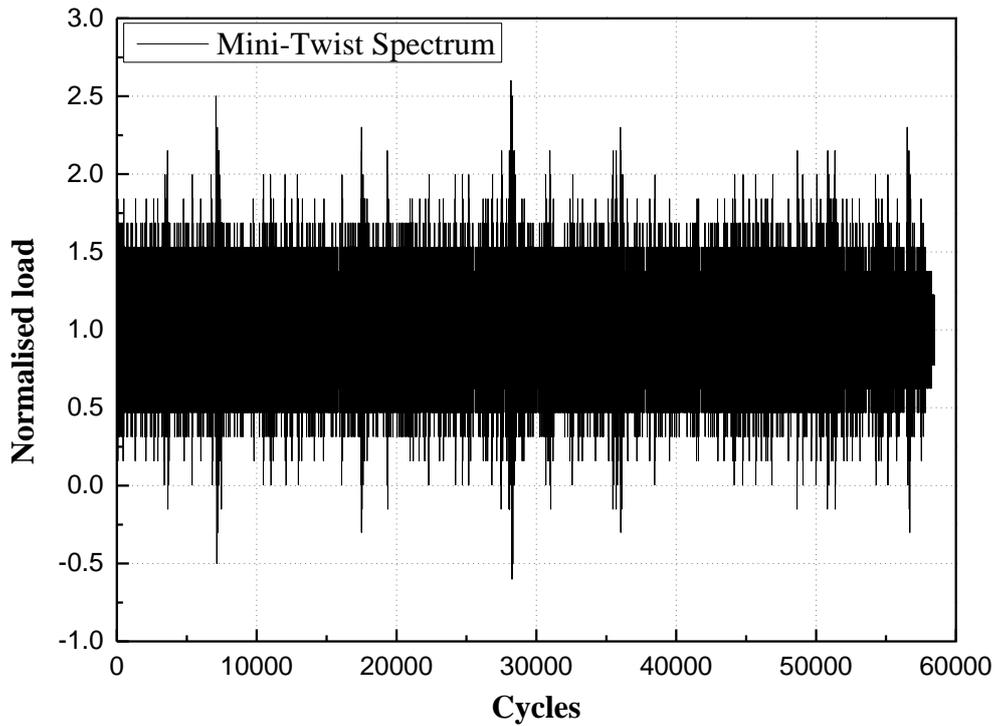

**Fig. 3** Random normalised load spectrum.

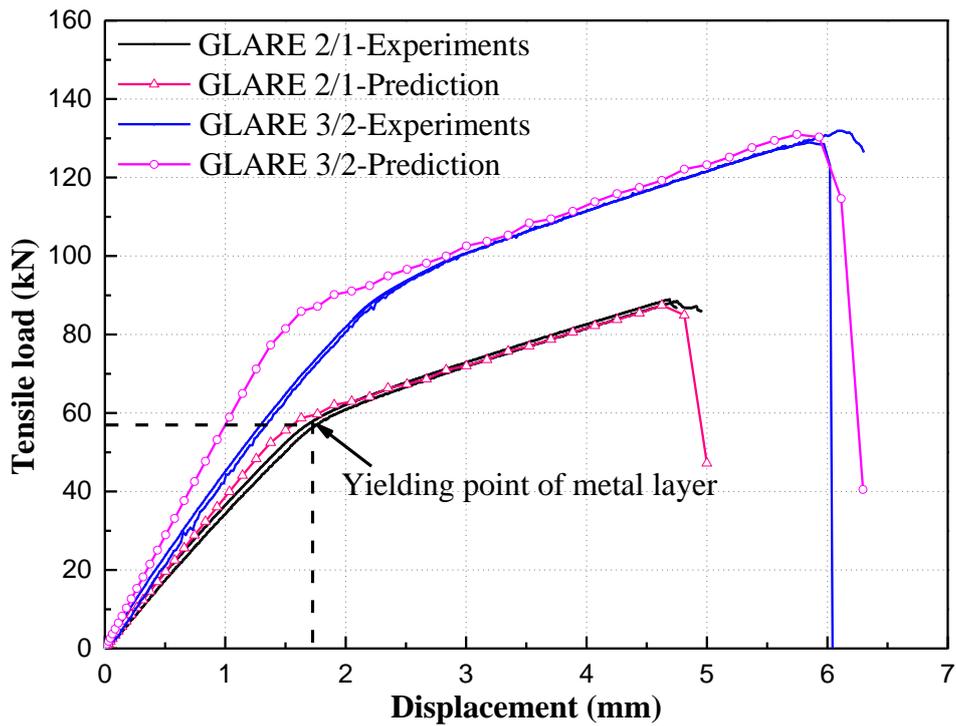

**Fig. 4** Tensile load versus displacement curves of the GLARE 2/1 and GLARE 3/2 laminates (Black and blue straight lines represent experiments for GLARE 2/1 and GLARE 3/2 laminates, respectively; red straight line with triangle symbol represents prediction of GLARE 2/1 laminates, while magenta straight line with circle symbol represents prediction of GLARE 3/2 laminates).

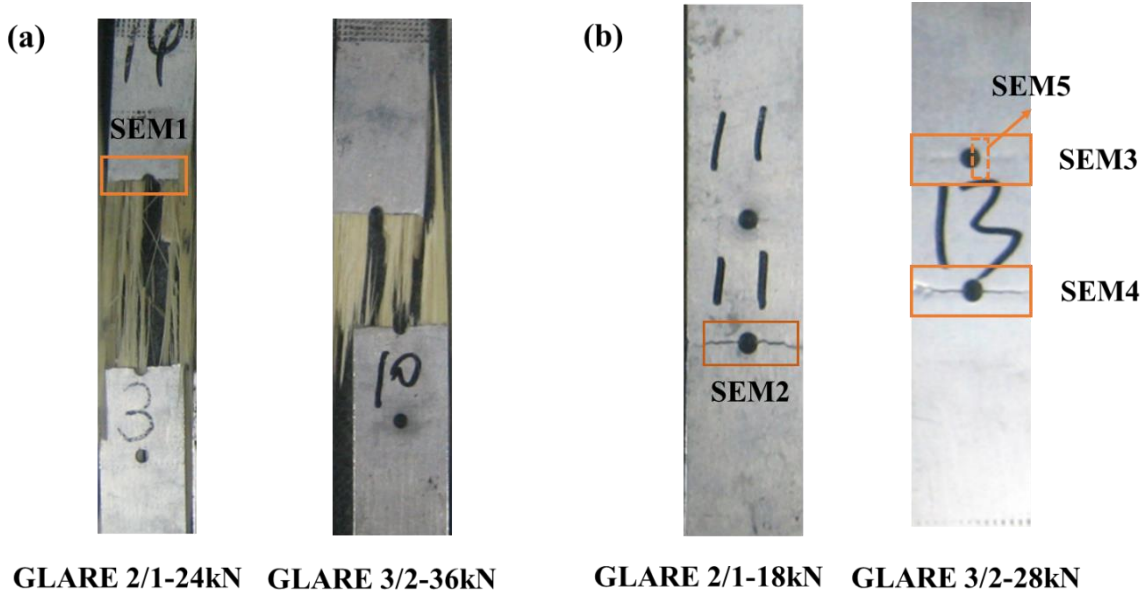

**Fig. 5** Typical macroscopic failure topologies: (a) Mode I: GLARE 2/1 and GLARE 3/2 laminates under random loading spectrum at reference load levels of 24 kN and 36 kN, respectively; (b) Mode II: GLARE 2/1 and GLARE 3/2 laminates under random loading spectrum at reference load levels of 18 kN and 28 kN, respectively.

Table 3  Ultimate tensile strengths and fatigue lifetimes under random loading spectrum.

| Specimen type | Ultimate tensile strength (kN) | Reference load level (kN) | Experiments | | Predictions | | | |
|---|---|---|---|---|---|---|---|---|
| | | | number of loading cycle (cycles) | Mean life (blocks) | without load sequence effect of metal layer | Relative deviation | with load sequence effect of metal layer | Relative deviation |
| GLARE 2/1 | 88.2 88.4 | 24 (High) | 46220,61914, 55242,76048 | 1.024 | 0.854 | 16.65% | 0.984 | 3.89% |
| | | 21 (Middle) | 108584,197652, 178955,125053 | 2.610 | 2.146 | 17.79% | 2.505 | 4.04% |
| | | 18 (Low) | 348877,262398, 273966,430206 | 5.627 | 3.983 | 29.23% | 4.505 | 19.94% |
| GLARE 3/2 | 128.9 131.2 | 36 (High) | 74364,79897, 64793,72971 | 1.249 | 0.878 | 29.74% | 0.983 | 21.35% |
| | | 33 (Middle) | 123424,124073, 115449,125603 | 2.090 | 1.505 | 27.99% | 1.867 | 10.65% |
| | | 28 (Low) | 247652,248301, 258893,261405 | 4.347 | 3.082 | 29.10% | 3.505 | 19.37% |

## 3.3 SEM analysis

To analyse fatigue failure mechanisms of the GLARE laminates in-depth, JEOL JSM-6010 scanning electron microscopy (SEM) was used to observe five sites of typical failure specimens for two types of the GLARE laminates (see Fig. 5). Fig. 6 and Fig.7 illustrate typical fracture morphologies of GLARE laminates under random loading spectrum. From Fig. 6, it is apparent that there are remarkable differences between fracture morphologies of the Al-Li alloy layers in the GLARE 2/1 laminates under random loading spectrum at low and high reference load levels. Alternatively,

a) At a low reference load level of 18 kN, fatigue crack initiation, growth and instant fracture regions can be observed on fracture morphologies of two Al-Li alloy layers (see Fig. 6(a)), and fracture surfaces of crack initiation and growth regions look like smooth and bright. Moreover, close fatigue stripes of two Al-Li alloy layers indicate that fatigue crack initiation, growth and fracture processes of all Al-Li alloy layers on the GLARE 2/1 laminates seem close. This agrees with the observation in previous work[40] that smooth region of fracture morphologies for all aluminium alloy layers on the GLARE laminate is almost the same. This is because the Al-Li alloy layer dominates fatigue behaviours of the GLARE laminates at low fatigue stress level, and glass fibre layers slow down crack growth rate after the cracks initiate first on one Al-Li alloy layer accompanied by crack initiation and growth on the other Al-Li alloy layers, resulting in similar fatigue failure process on all Al-Li alloy layers. In addition, many obvious retardation lines present on crack growth region because of the crack retardation caused by multiple tensile overloads under random loading spectrum (see Figs. 6(a)-6(d)), illustrating that load sequence has a significant effect on fatigue failure mechanisms of the Al-Li alloy layer in the GLARE laminates under random loading spectrum, which should be considered in fatigue life prediction model.

b) At a high reference load level of 24 kN, fracture morphologies of the Al-Li alloy layers are dark and rough, and the locations of instant fracture regions vary dramatically (see Figs. 6(e) and 6(f)). The reason for this is that glass fibre layers govern fatigue behaviours of GLARE laminates at high fatigue stress level. The Al-Li alloy layers lose fibre bridging in the next fatigue cycles after glass fibre layers fail, leading to the quick breaking of the Al-Li alloy layers at different sites.

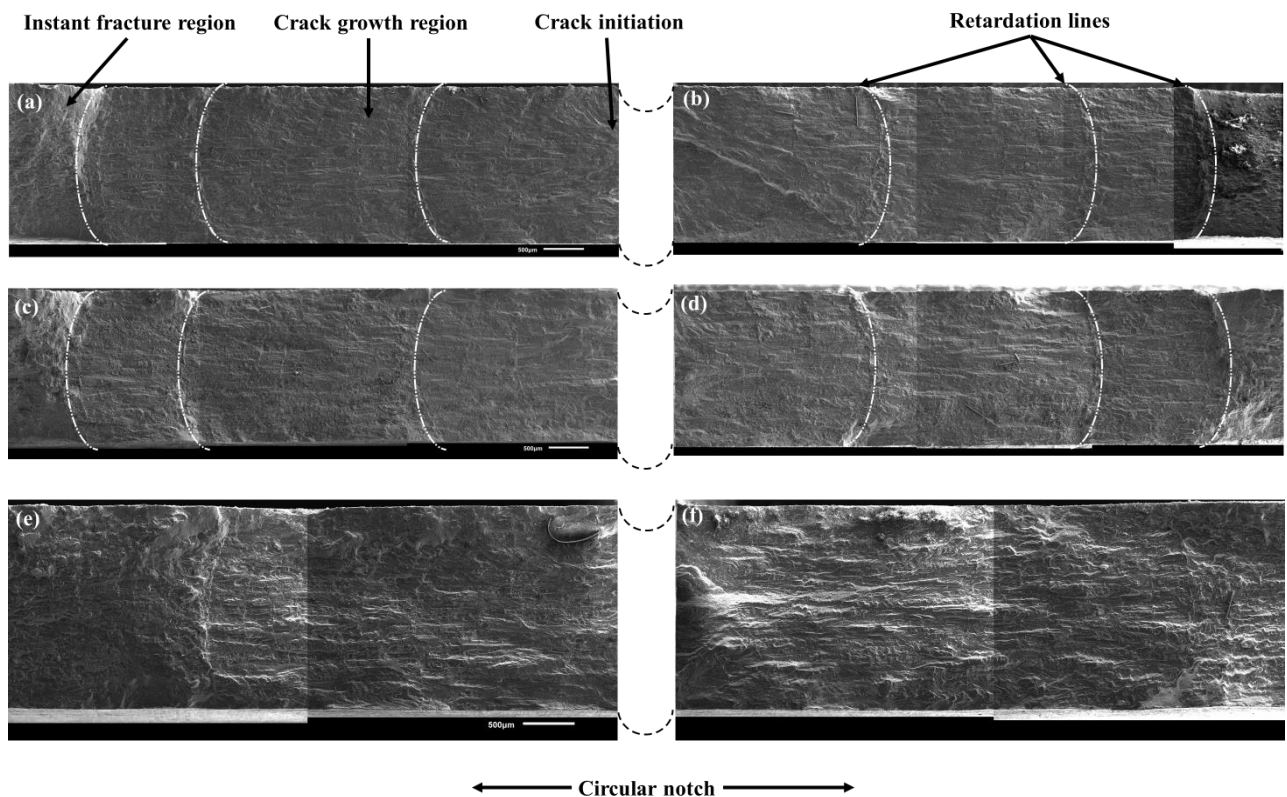

**Fig. 6** Fracture morphologies of the Al-Li alloy layers in the GLARE 2/1 laminate under random loading spectrum: (a)-(d) SEM 2, both left and right cross sections of circular notch on two Al-Li alloy layers at a low reference load level of 18 kN; (e)-(f) SEM 1, both left and right cross sections of circular notch on an Al-Li alloy layer at a high reference load level of 24 kN.

From Fig. 7, it is clear that fracture morphologies of the GLARE 3/2 laminates at a low reference load level of 28 kN under random loading spectrum are similar to those of the GLARE 2/1 laminates at a low reference load level of 18 kN (see Figs. 7(a)-7(h)). Fracture morphologies of three Al-Li alloy layers on the GLARE 3/2 laminates have also obvious crack initiation, growth and instant fracture regions, and fracture surfaces of crack initiation and growth regions are also smooth and bright. Meanwhile, there are also multiple clear retardation lines on crack growth region. However, the fracture of the Al-Li alloy layer, the breakage of glass fibre, and delamination of fibre-metal matrix layer can be observed at the local zone around circular notch on fracture morphology of the GLARE 3/2 laminates (see Figs. 7(i) and 7(j)), which was also found in literature [6]. This result implies that fatigue failure of the GLARE laminate is dominated by the interactive fatigue failure mechanisms of three-phase materials, which should be also involved in fatigue life prediction model.

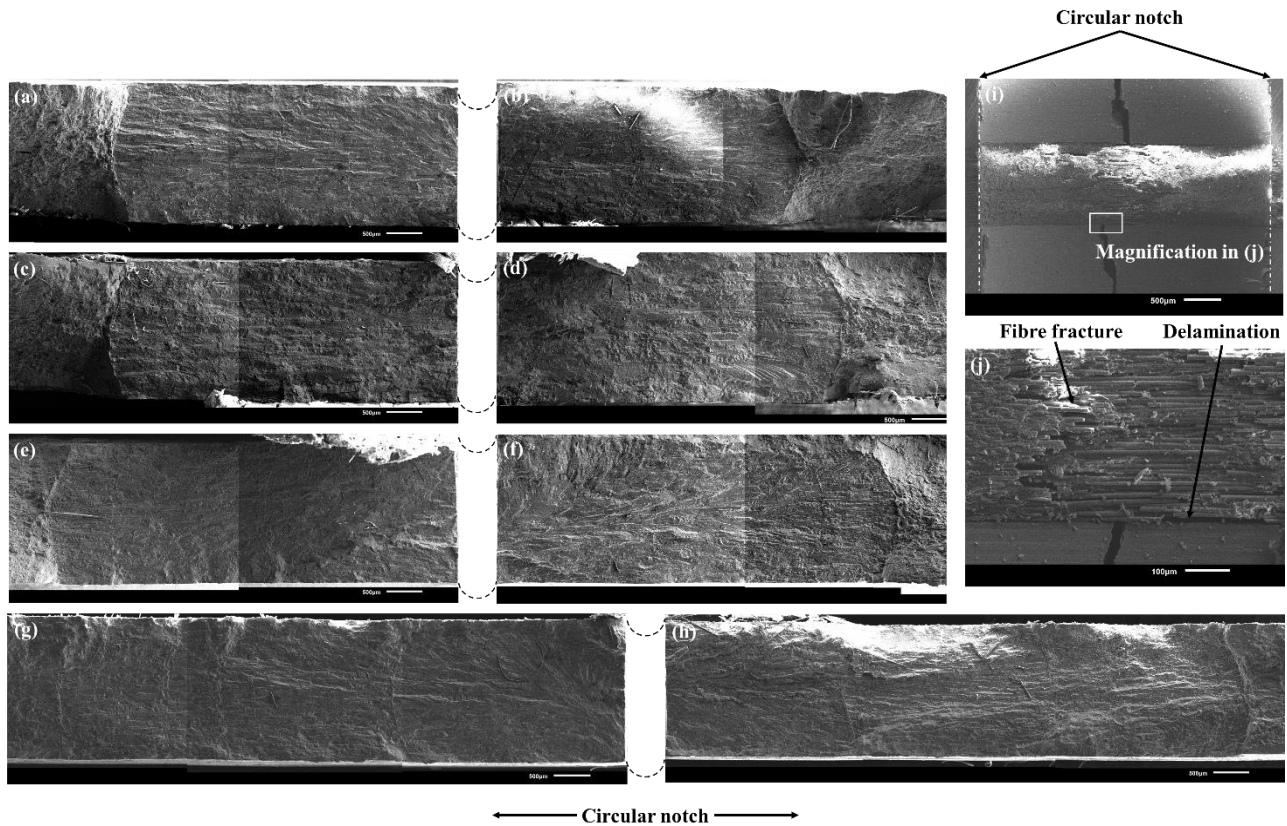

**Fig. 7** Fracture morphologies of the GLARE 3/2 laminate under random loading spectrum at a low reference load of 28 kN: (a)-(f) SEM 3, partial width failure on both left and right cross sections of circular notch on three Al-Li alloy layers; (g)-(h) SEM 4, full width failure on both left and right cross sections of circular notch on an Al-Li alloy layer; (i) SEM 5, partial area around circular notch; (j) magnification of (i).

## 4 Validation

### 4.1 FE model

Due to the geometric symmetry of the GLARE 2/1 and GLARE 3/2 laminate specimens as shown in Fig. 2, it is reasonable and computationally efficient to build a symmetrical 1/8 three-dimensional FE model in length, width and thickness directions of the specimen. The local coordinates are set up to ensure three axial directions $x$, $y$ and $z$ of the coordinate system consistent with the longitudinal, transverse and through-thickness directions for the specimens (see Fig. 8). Reduced integration 8-node linear brick solid elements (C3D8R) are used to model glass fibre and Al-Li alloy layers. To eliminate the effect of element size as possible, four element sizes of 1.0 mm, 0.5 mm, 0.4 mm and 0.3 mm in the near-circular notch zone and the element size of 2.0 mm in the far-field region are selected to implement preliminary static simulations, and the results have been converged at element sizes of 0.3 mm in the near-circular notch zone and 2.0 mm in the far-field region. Therefore, the element mesh sizes of about 0.3 mm, 1.0 mm and 2.0 mm are used in the near-circular notch zone, transition zone and far-field zone, respectively, for obtaining reliable numerical results at a relatively low computational cost. To availably model delamination behaviour of matrix layer on fibre-metal interfaces, cohesive element COH3D8 is embedded between glass fibre and Al-Li alloy layers with a thickness of 0.001 mm. The FE model of the GLARE 2/1 laminate has 12640 C3D8R elements and 1264 COH3D8 elements (see Fig. 8(a)), and that of the GLARE 3/2 laminate has 15904 C3D8R elements and 2272 COH3D8 elements (see Fig. 8(b)). Symmetric constraints are assigned into three symmetric planes of the FE model. The far end face of the FE model is firstly coupled to the reference point RP1, and then 1/4 quasi-static tensile loading is applied to this reference point. General contact is employed with the contact properties of hard contact in the normal direction and penalty stiffness contact with the friction coefficient of 0.2 in the tangential direction.

Progressive static damage algorithm of glass fibre layer is written in the VUMAT subroutine, which composes of constitutive model, static failure criteria and sudden stiffness degradation rule (see Eq. (2)). Static failure criteria are in the same form as fatigue failure criteria listed in Table 1, but it does not experience fatigue cyclic loading, so the term of strength reduction in fatigue failure criteria is zero. Mechanical properties of S4/SY-14 glass fibre lamina are listed in Table 2. Isotropic hardening constitutive model and ductile fracture model with linear degradation rule are used to characterise elastic-plastic mechanical behaviours and damage failure behaviours of the Al-Li alloy layer in the GLARE laminate, and relevant material properties of 2060 Al-Li alloy sheet are listed in Table 2. Delamination behaviours of matrix layer on fibre-metal interface in the GLARE laminate are characterised by the CZM, and relevant model parameters used in this work are listed in Table 4. The Abaqus/Explicit is used for all numerical analysis in this paper, and mass-scaling control is set with the target time increment of $2.5 \times 10^{-5}$ s to obtain reliable numerical results at a relatively low computational cost.

The predicted tensile load versus displacement curves of the GLARE 2/1 and GLARE 3/2 laminates are shown in Fig. 4. It is evident from Fig. 4 that the predicted tensile load versus displacement curves are in good agreement with the experimental results. Besides, the predicted ultimate tensile loads of the GLARE 2/1 and GLARE 3/2 laminates are respectively 87.44 kN and 130.98 kN, and the experimental mean values of those are separately 88.30 kN and 130.05 kN (see Table 3), so the relative deviations are within 2%. Consequently, the above results confirm that the symmetric 1/8 three-dimensional FE model is accurate and effective enough.

**Table 4**  Model parameters of the progressive fatigue delamination damage model[41][42].

| $k_I$ (GPa/mm) | $k_{II}$ (GPa/mm) | $X_I$ (MPa) | $X_{II}$ (MPa) | $G_{IC}$ (mJ/mm$^2$) |
|---|---|---|---|---|
| 475 | 175 | 49.8 | 73.7 | 0.52 |
| $G_{IIC}$ (mJ/mm$^2$) | $\eta$ | $C_3$ | $m_3$ | / |
| 1.61 | 1.89 | 0.005 | 0.75 | / |

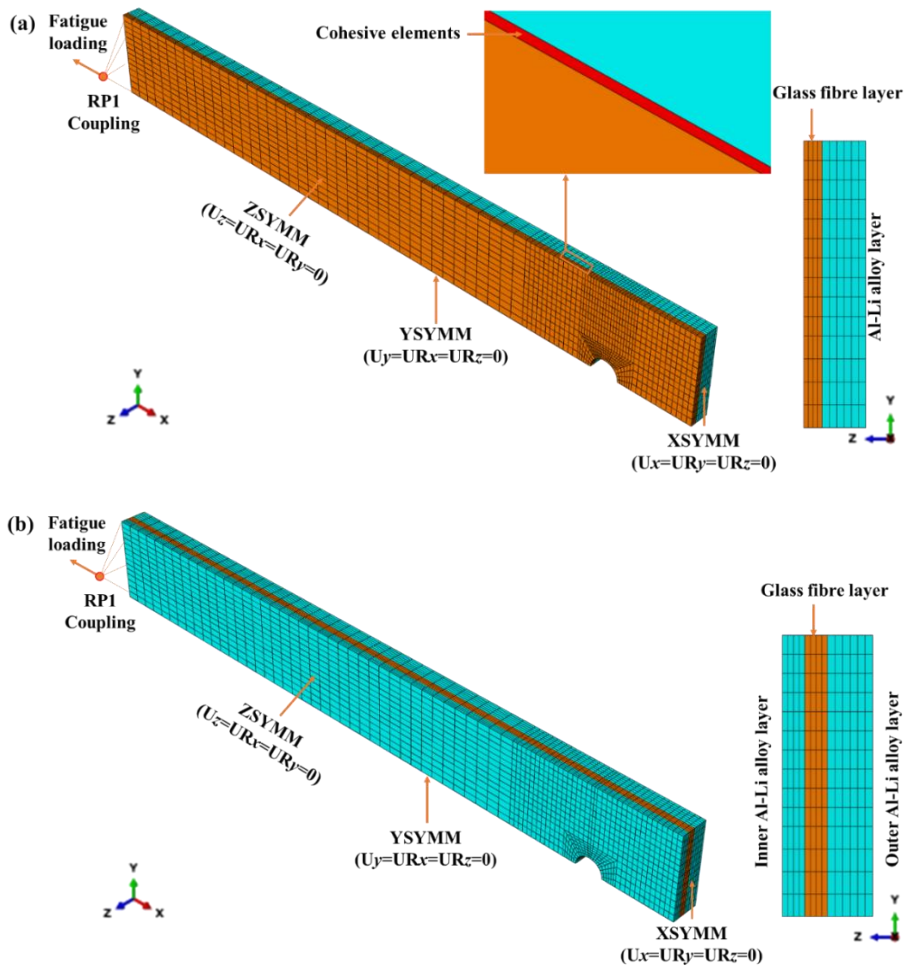

**Fig. 8**  FE models: (a) GLARE 2/1 laminate; (b) GLARE 3/2 laminate.

## 4.2 Mixed algorithm

The boundary conditions of fatigue FE model are the same as that of static FE model in Section 4.1. Note that, each fatigue loading cycle is modelled into the quasi-static loading with the same magnitude as the maximum absolute value of fatigue loading cycle. All fatigue loading cycles are extracted from the random loading spectrum in Fig. 3 and the sequence number for each loading cycle is recorded by utilising the rain-flow counting method. To simulate fatigue loading history under random loading spectrum, the 1/4 reference load level is first applied to the reference point RP1, and then load coefficient history processed by the rain-flow counting method is assigned to the smooth loading amplitude curve. According to the mixed progressive fatigue damage algorithm of three-phase materials shown in Fig. 1, progressive fatigue damage analysis of the GLARE laminate subjected to random loading spectrum is carried out in the Abaqus/Explicit code.

In general, static tests and uniaxial tension-tension and compression-compression constant amplitude fatigue tests in longitudinal, transverse, in-plane shear, and out-plane shear directions of S4/SY-14 glass fibre lamina are necessary to determine all model parameters in Eq. (1). In this work, owing to the characteristics of unidirectional longitudinal glass fibre layers and tension-dominated fatigue loading conditions under random loading spectrum, the significant strength and stiffness degradation behaviours of fibre layer occur in the longitudinal tension direction, while those degradation behaviours can be negligible in other material principal directions. Hence, it is practical and reasonable to obtain longitudinal tension fatigue properties of glass fibre lamina only, and the relevant parameters are listed in Table 5. Besides, it is essential to conduct constant amplitude fatigue and crack growth tests for monolithic metal corresponding to metal layer in the FML to determine all model parameters in Eqs. (13) and (14). However, due to the limitation of resource, constant amplitude fatigue test data of 2060 Al-Li alloy sheet in previous literature[10][43][44] are adopted to determine the model parameters in Eq. (13). Fatigue properties of the monolithic 2060 Al-Li alloy sheet corresponding to metal layer in the GLARE laminates used in this paper are listed in Table 5.

**Table 5** Fatigue properties parameters of S4/SY-14 glass fibre lamina and 2060 Al-Li alloy sheet.

| S4/SY-14 glass fibre lamina | $r_0$ | $H_{1t}$ | $p_{1t}$ | $q_{1t}$ | $S_{0,1t}$ | $X_{1t}$ |
|---|---|---|---|---|---|---|
| | 0.06 | $2.457 \times 10^7$ | 1.33 | -2.98 | 709.83 | 2000 |
| 2060 Al-Li alloy sheet[10],[28][43][44] | $r_0$ | $\alpha$ | $\beta$ | $S_0$ | $C$ | $C_1$ |
| | 0.06 | -0.493 | -2.427 | 170 | $1.584 \times 10^8$ | $1.580 \times 10^{-8}$ |
| | $m$ | $m_1$ | $m_2$ | $\Delta K_{th}$ | $\sigma_s$ | $r_{so}$ |
| | 1.73 | 3.29 | -1.71 | 0.52 | 470 | 2.30 |

Fatigue lives of random loading spectrum predicted by the mixed progressive fatigue damage algorithm of three-phase materials are listed in Table 3. The simulated fatigue behaviours of the

GLARE 3/2 laminates under random loading spectrum at high and low reference load levels are depicted in Figs. 9 and 10 and Table 6. From Fig. 9 and Table 6, fatigue behaviours of the GLARE 3/2 laminates under random loading spectrum at a high reference load level of 36 kN can be illustrated as follows:

a) In the first 1400 cycles, outer and inner Al-Li alloy layers bear main fatigue loading because the stiffness of the Al-Li alloy layers is greater than that of glass fibre layers, and fatigue damage of outer and inner Al-Li alloy layers firstly appears at stress concentration sites around circular notch. Besides, a small amount of delamination is found in two matrix layers on fibre-metal interfaces due to the stiffness mismatch between fibre and metal layers. There is no fibre and matrix damage in glass fibre layers.

b) At 8516 cycles, plastic stress flow in outer and inner Al-Li alloy layers causes stress redistribution of each layer in the GLARE laminate, which results in glass fibre layers carrying predominant fatigue loading. Delamination in two matrix layers on fibre-metal interfaces grows further. Although no fibre and matrix failures occur in glass fibre layers, residual strength of glass fibre layers declines irreversibly.

c) In the following six cycles, i.e. 8522 cycles, obvious delamination growth happens in two matrix layers on fibre-metal interfaces, and then a little fibre and matrix failures germinate in glass fibre layers that primarily bear the loads. Meantime, fatigue damage of outer and inner Al-Li alloy layers accumulates slowly.

d) At 28589 cycles, delamination grows slightly in two matrix layers on fibre-metal interfaces, and fibre and matrix failures propagate further in glass fibre layers. Moreover, fatigue damage increases slowly in outer and inner Al-Li alloy layers.

e) In the 56016 cycles, two matrix layers on fibre-metal interfaces almost completely fail, and fibre and matrix failures of glass fibre layers grow dramatically. The cumulative fatigue damage of outer and inner Al-Li alloy layers increases moderately.

f) In the final 57422 cycles, two matrix layers on fibre-metal interfaces fail completely, and the fibre and matrix of glass fibre layers fail on whole cross-section of specimen, resulting in the bearing capacity loss of glass fibre layers. The outer and inner Al-Li alloy layers passively carry loads alone, and instant tensile fracture occurs because the external loads exceed their ultimate strengths.

Again, from Fig. 10 and Table 6, fatigue behaviours of the GLARE 3/2 laminates under random

loading spectrum at a low reference load level of 28 kN can be outlined as follows:

a) In the initial 4806 cycles, fatigue damage appears first at stress concentration sites around circular notch in outer and inner Al-Li alloy layers. There is a small amount of delamination in two matrix layers on fibre-metal interfaces. Besides, no fibre and matrix failures are found in glass fibre layers.

b) At 8522 cycles, fatigue damage of outer and inner Al-Li alloy layers and delamination of two matrix layers on fibre-metal interfaces grow successively, while fibre layers remain intact.

c) In the 146387 cycles, both fatigue damage of outer and inner Al-Li alloy layers and delamination of two matrix layers on fibre-metal interfaces propagate significantly. A little matrix failure germinates in glass fibre layers, but fibre layer still remains intact.

d) In the next seven cycles, i.e. 146394 cycles, fatigue damage and delamination increase slightly in outer and inner Al-Li alloy layers and two matrix layers on fibre-metal interfaces, respectively. In addition, matrix failure grows mildly and little fibre failure happens for the first time in glass fibre layers.

e) At 204835 cycles, delamination grows dramatically in two matrix layers on fibre-metal interfaces. Fatigue damage grows to more than half specimen width in outer and inner Al-Li alloy layers. However, fibre and matrix failures increase slightly in glass fibre layers.

f) In the final 204841 cycles, two matrix layers on fibre-metal interfaces delaminate thoroughly, and outer and inner Al-Li alloy layers fracture along the specimen width direction. Glass fibre layers then bear loads alone and fracture rapidly, accompanying a large area of tensile fibre and matrix failures.

Fatigue behaviours of the GLARE 2/1 are similar to those of the GLARE 3/2 laminates under random loading spectrum. It is clear that fatigue failure of the GLARE laminate depends on the reference load level of random loading spectrum. Specifically,

a) At a high reference load level, predominant fatigue failure mechanism of the GLARE laminate is fatigue failure of glass fibre layer, and delamination growth in fibre-metal interface matrix layer is prior to fibre or matrix failure growth in glass fibre layer. After fibre-metal interface matrix layer delaminates completely, glass fibre layer breaks, followed by the rapid fracture of the Al-Li alloy layer, which is likely to lead to Mode I failure of the GLARE laminate.

b) At a low reference load level, predominant fatigue failure mechanism of the GLARE laminate is fatigue fracture of the Al-Li alloy layer, and delamination in matrix layer on fibre-metal interface grows before fatigue damage in the Al-Li alloy layer propagates. After fibre-metal interface matrix layer fails thoroughly, the Al-Li alloy layer fractures along the specimen width direction and then glass fibre layer breaks. This is possible to result in Mode II failure of the GLARE laminate.

The above results correlate well with macroscopic and microscopic failure mechanism observed in Sections 3.2 and 3.3.

It is clear from Table 5 that maximum relative deviations between fatigue life predictions and experiments by using the proposed mixed algorithm without and with considering the load sequence effect of metal layer are 29.74% and 21.35%, respectively, which indicates that the proposed mixed algorithm with considering the load sequence effect of metal layer can obtain more accurate numerical results. Moreover, the predicted fatigue failure behaviours of the GLARE laminates under random loading spectrum by using the proposed mixed algorithm agree well with experimental results. The above results demonstrate that the proposed mixed algorithm can effectively simulate fatigue behaviours and lives of the GLARE laminate under random loading spectrum.

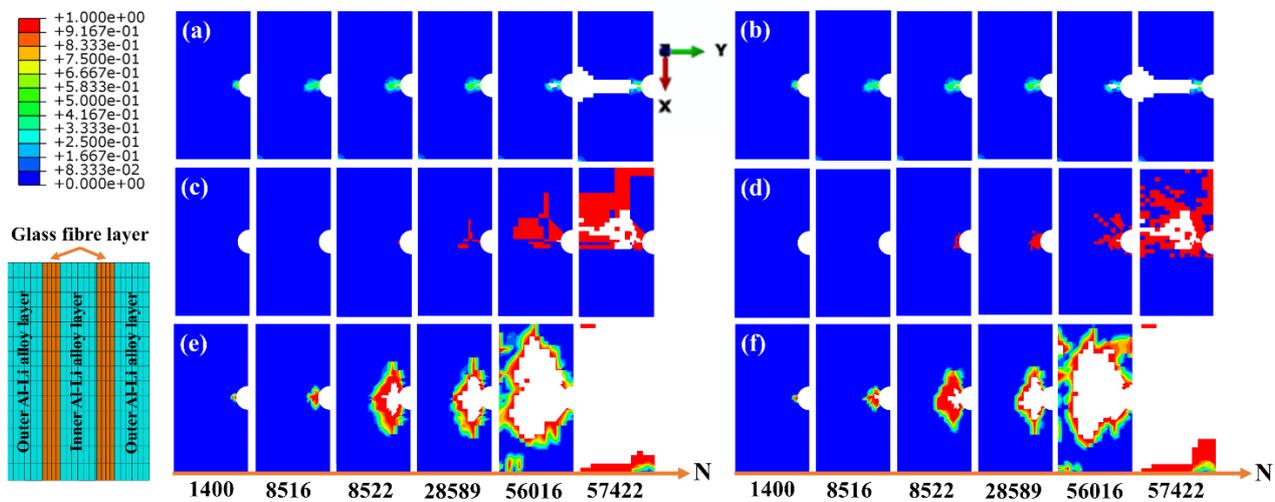

**Fig. 9** Predicted fatigue failure behaviours of the GLARE 3/2 laminate under random loading spectrum at a high reference load of 36 kN: (a) Fatigue damage of outer Al-Li alloy layer; (b) Fatigue damage of inner Al-Li alloy layer; (c) Fibre damage of glass fibre layer; (d) Matrix damage of glass fibre layer; (e) Delamination of outer fibre-metal interface matrix layer; (f) Delamination of inner fibre-metal interface matrix layer.

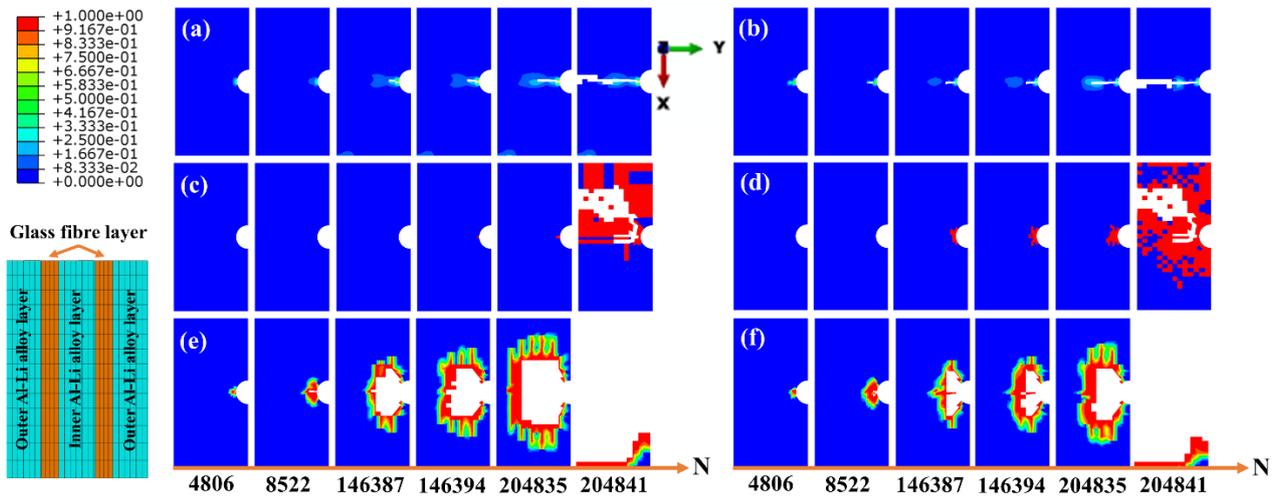

**Fig. 10** Predicted fatigue failure behaviours of the GLARE 3/2 laminate under random loading spectrum at a low reference load of 28 kN: (a) Fatigue damage of outer Al-Li alloy layer; (b) Fatigue damage of inner Al-Li alloy layer; (c) Fibre damage of glass fibre layer; (d) Matrix damage of glass fibre layer; (e) Delamination of outer fibre-metal interface matrix layer; (f) Delamination of inner fibre-metal interface matrix layer.

**Table 6** Predicted fatigue failure behaviours of the GLARE 3/2 laminate under random loading spectrum.

| Reference load level (kN) | Fatigue failure behaviours of the GLARE 3/2 laminate under random loading spectrum |
|---|---|
| 36 (High) | **a)** at 1400 cycles, fatigue damage of outer and inner Al-Li alloy layers firstly appears around circular notch; a small amount of delamination is found in two fibre-metal interfaces; there is no fibre and matrix damage in glass fibre layers.<br>**b)** at 8516 cycles, glass fibre layers carrying predominant fatigue loading; delamination in two fibre-metal interfaces grows further; residual strength of glass fibre layers declines irreversibly.<br>**c)** at 8522 cycles, obvious delamination growth happens in two fibre-metal interfaces; a little fibre and matrix failures germinate in glass fibre layers; fatigue damage of outer and inner Al-Li alloy layers accumulates slowly.<br>**d)** at 28589 cycles, delamination grows slightly in two fibre-metal interfaces; fibre and matrix failures propagate further in glass fibre layers; fatigue damage increases slowly in outer and inner Al-Li alloy layers.<br>**e)** at 56016 cycles, two fibre-metal interfaces almost completely fail; fibre and matrix failures of glass fibre layers grow dramatically; the cumulative fatigue damage of outer and inner Al-Li alloy layers increases moderately.<br>**f)** at 57422 cycles, two fibre-metal interfaces fail completely; fibre and matrix of glass fibre layers fail on whole cross-section of specimen; the outer and inner Al-Li alloy layers fracture instantly. |
| 28 (Low) | **a)** at 4806 cycles, fatigue damage appears first around circular notch in outer and inner Al-Li alloy layers; there is a small amount of delamination in two fibre-metal interfaces; no fibre and matrix failures are found in glass fibre layers.<br>**b)** at 8522 cycles, fatigue damage of outer and inner Al-Li alloy layers and delamination in two fibre-metal interfaces grow successively, while fibre layers remain intact.<br>**c)** at 146387 cycles, fatigue damage of outer and inner Al-Li alloy layers and delamination in two fibre-metal interfaces propagate significantly; a little matrix failure germinates in glass fibre layers, but fibre layer still remains |

|  | intact. |
|--|---------|
|  | **d)** at 146394 cycles, fatigue damage and delamination increase slightly in outer and inner Al-Li alloy layers and fibre-metal interfaces, respectively; matrix failure grows mildly and little fibre failure happens for the first time in glass fibre layers. |
|  | **e)** at 204835 cycles, delamination grows dramatically in two fibre-metal interfaces; fatigue damage grows to more than half specimen width in outer and inner Al-Li alloy layers; fibre and matrix failures increase slightly in glass fibre layers. |
|  | **f)** at 204841 cycles, two fibre-metal interfaces delaminate thoroughly; outer and inner Al-Li alloy layers fracture; glass fibre layers then bear loads alone and fracture rapidly, accompanying a large area of tensile fibre and matrix failures. |

## 5 Conclusions

This paper presents an experimental and numerical study on fatigue failure behaviours of novel GLARE laminates made of S4/SY-14 glass fibre prepreg and 2060 Al-Li alloy sheet under random loading spectrum. The proposed mixed algorithm based on fatigue damage concepts of three-phase materials and revealed fatigue failure mechanisms of the FML under random loading spectrum are the main novel contributions of this work. The following conclusions can be drawn from this investigation:

**(i)** A mixed algorithm based on fatigue damage concepts of three-phase materials (including fibre layer, metal layer, and fibre-metal interface matrix layer) is proposed for modelling progressive fatigue damage mechanisms and fatigue life of the FML under random loading spectrum.

**(ii)** Predominant fatigue failure of the GLARE laminate depends on the reference load level of the random loading spectrum. That is, dominant fatigue failure of the GLARE laminate is dependent on fatigue strength of fibre layer at a high reference load level, but metal layer at a low reference load level. In addition, fatigue delamination growth in fibre-metal interface matrix layer is always prior to fibre and matrix failure growth in fibre layer or fatigue damage propagation in metal layer regardless of the reference load level.

**(iii)** Fatigue fracture morphologies of the Al-Li alloy layers in the GLARE laminates vary with the reference load level of random loading spectrum. At a low reference load level, there is distinct crack initiation, growth and instant fracture regions. Fracture surfaces of crack initiation and growth regions are smooth and bright, and many clear retardation lines present on crack growth region because of the crack retardation caused by multiple tensile overloads under random loading spectrum. However, fracture surfaces of Al-Li alloy layers are dark and rough, and the locations of instant fracture regions are different remarkably due to the unstable and quick fracture of metal layers after the failure of fibre layers at a high reference load level.

**(iv)** A symmetrical 1/8 FE model is built and validated by static tension tests. Based on the verified FE model and material properties of constituents, the proposed mixed modelling algorithm based on fatigue damage concepts of three-phase materials has been employed to predict fatigue failure behaviours and lives of the GLARE laminates under random loading spectrum. A good correlation is achieved between predictions and experiments, demonstrating the effectiveness and accuracy of the proposed modelling algorithm.

**Various possibilities can be envisaged to continue this investigation:**

**(1)** The quantitative effect of variable frequency on the strength and stiffness behaviours of the fibre layer, fatigue strength and crack growth of the metal layer, and fatigue delamination growth of the fibre-metal interface needs to be considered in the fatigue models of three-phase materials for modelling the effect of variable-frequency random-loadings in the future.

**(2)** It seems necessary for more fatigue test results of fibre metal laminates to further validate the proposed mixed algorithm. For example, block loading fatigue tests should be implemented under two-stage low-high or high-low sequences to demonstrate explicitly the mixed algorithm's capability of considering the load sequence effect.

## Acknowledgements

This project was supported by the National Natural Science Foundation of China (Grant No. 51875021) and the China Scholarship Council (Grant No. 202006020210). W. Tan acknowledges financial support from the EPSRC, United Kingdom (Grant EP/V049259/1).